\documentclass[referee]{aa}
\usepackage{graphics}
\begin{document}
\thesaurus{04(11.13.2; 11.19.2; 11.19.6)}
\title{The truncation of stellar discs. A theoretical model}

\author{E. Battaner, E. Florido  \and J. Jimenez-Vicente }
\institute{Dpto. Fisica Teorica y del Cosmos. Universidad de
Granada. Spain}
\titlerunning{The truncation of stellar discs}
\authorrunning{Battaner et al}
\maketitle
\begin{abstract}

The truncation of stellar discs is not abrupt 
but characterized by a continuous distancing from the exponential
profile. There exists a truncation curve, $t(r)$, ending at a
truncation radius, $r_t$. We present here a theoretical model in which
it is assumed that the magnetic hypothesis explaining the flat
rotation curve also explains the truncation. Once stars are born, the
centripetal magnetic force previously acting on the progenitor gas cloud
is suddenly interrupted, and stars must move to larger orbits or
escape. The agreement between theoretical and observed truncation
curves is very satisfactory. Parameters defining the disc gas rotation
curve should therefore be related to those defining the
truncation. It is predicted that rotation curves that quickly reach the
asymptotic value $\theta_0 = \theta (r=\infty)$ would have small truncation
radii. On the contrary, $r_t$ and $\theta_0$ itself, would be
uncorrelated quantities.

\end{abstract}

\section{Introduction}

The interest in studying the truncation of stellar discs lies in the
fact that this is a phenomenon present in all spirals and that no
theoretical model has yet been advanced.

The truncation of stellar discs was discovered by van der Kruit (1979)
 and was the object of a preliminary but noticeably precise description by
van der Kruit \& Searle (1981a,b; 1982a,b) by means of photographic
photometry. Some basic facts were established in these pioneer works and
are noted here: a) as the involved intensities are very low,
truncations are better observed in edge-on galaxies; b) the truncation
radius, $r_t$, is about 4.2 times the radial scale length, $R$; c)
the truncation is not a sharp cut-off, but the radial e-folding drops
to about 1 kpc; therefore, there is a truncation curve, $t(r)$, with
$t(r_t)= \infty$, which will be precisely defined later.

It is evident that this phenomenon reveals important dynamic
effects, particularly if it is as common as it seems. More than a decade
after the work by van der Kruit \& Searle, the subject was
reconsidered by Barteldrees \& Dettmar (1994) who have renewed the
interest in this topic. In this CCD photometry study, it was
proposed that $r_t/R <3$, noticeably lower than the corresponding
value obtained by van der Kruit \& Searle.

Studies by Hamabe (1982), Sasaki (1987) and Hamabe \&
Wakamatsu (1989) have also considered truncation. In our Galaxy, Habing (1988)
found $r_t$ to be 9.5 kpc; Robin et al. (1992), 14 kpc; Ruphy et al.
(1996), 15 kpc; Porcel et al. (1997) found
$r_t \le$15 kpc, assuming the truncation interpretation for the near
infrared COBE data, following a discussion by Freudenreich et al.
(1994).

The truncation of the stellar disc is a common feature of all spiral
galaxies. For instance, van der Kruit \& Searle (1982a) detected it
in the four galaxies studied, as did Barteldrees \& Dettmar in  a
sample of 27 edge-on galaxies. Although it must be observed 
in noisy conditions, it is clear that it
is a universal phenomenon. The large sample of galaxies in nearby
clusters by Gavazzi et al. (1990, 1994, 1995) and
Gavazzi \& Randone (1994) 
gives an approximate truncation frequency of around 0.6, considering
just  edge-on non interacting galaxies.

Therefore, even if the relatively sharp truncation takes place at very
low surface brightness (greater than about 24 mag arcsec$^{-2}$), 
it is clear that it is a
universal phenomenon. Hence, it is highly interesting. This interest
is in contrast with the scarce number of statistical studies of
truncation reported in the literature. Considering that it is a feature
related to dominant dynamic effects at the periphery of spiral
galaxies, it is also remarkable that such limited attention has been paid by
theoretical studies to explain it.

Previous theoretical hypotheses concerning truncation have been summarized
by de Grijs (1997). Fall \& Efstathiou (1980) suggested that
truncation takes place at those radii at which shear by differential
rotation overcomes selfgravity, so that gravitational collapse and
star formation are inhibited, but this idea did not provide
agreement with real truncation radii (van der Kruit \& Searle,
1982a). Larson (1976) considered slow disc formation, so that the
truncation radius would just reflect the present age of the
galaxy. These early hypotheses were not developed with theoretical
models. Recently, Bottema (1996) proposed tides in interacting
galaxies as a cause of truncation, which could explain some but not
all the observed truncated discs. It can therefore be stated that, at
present, no compelling theory exists to explain this important dynamic 
phenomenon.

The magnetic hypothesis of the rotation curves (Nelson, 1988; Battaner
et al., 1992; Florido \& Battaner, 1995) provides a very straightforward
explanation of the truncation discs: the outer disc has no star
because they escape once they are formed. Under the magnetic
hypothesis, the rotating gas is subject to the centrifugal force in
equilibrium with two centripetal forces: gravitational and
magnetic. When gas forms stars, the centrifugal and the gravitational
forces remain the same but the magnetic suddenly disappears. Then the
star migrates to another orbit with a larger radius or even escapes.

The purpose of this paper is therefore to quantify this idea and to show
how real truncation curves are reproduced by the magnetic model.

\section{An analytical model}

We first need a rotation curve for the gaseous disc becoming flat at
large radii, and  another curve for the stellar disc, coincident
with the gas curve for small radii and Keplerian for large radii. We
then propose

\begin{equation}
\varphi_{gas} = {\theta_{gas} \over \theta_0}
\end{equation}

\begin{equation}
\varphi_{stars}= {\theta_{stars} \over \theta_0}
\end{equation}

\begin{equation}
x= {r \over R_F}
\end{equation}

\begin{equation}
\varphi_{gas}=1-e^{-x}
\end{equation}

\begin{equation}
\varphi_{stars}= \left( 1-e^{-x} \right)e^{-\left( {R_F \over R_K}
		\right)^2 x^2}+ 
 		Cx^{-{1 \over 2}} \left( 1-e^{-\left({R_F \over R_K}
		\right)^2 x^2}\right)
\end{equation}

where $R_F$ is a constant, a typical radial length which indicates how
slowly the gas rotation curve becomes flat; $R_K$ is another constant,
a typical radial length indicating where the stellar rotation curve
becomes Keplerian; $C$ is another constant, providing information on the
point central mass once $\varphi_{stars}$ become Keplerian.

We therefore assume a corotation region for $r \ll R_F$. For $r >
R_K$ the star rotation curve is Keplerian and the gas rotation curve
is flat. As we are working under the magnetic hypothesis we do not
need the presence of a dark halo. The Keplerian region would specify
the velocity of true stars in a steady orbit at radius $r$, if stars
really existed in this region. However, we will see that this
region is devoid of stars with stationary orbits, at least for a
large range of the parameters involved.

Neither the disc gas rotation curve nor the disc stellar rotation curve are
directly observable. In the innermost region the bulge dominates both
curves, and we only consider the disc component. In the
Keplerian region, star velocities are unobserved, because either the
luminosity is too low to be appreciated with present techniques, or
because of the complete absence of stars. However, observations very
much restrict our choices of both curves. The parameter $\theta_0$ is
observational and is known for most flat rotation galaxies; $R_F$ is
more or less related to the radius at which $\theta$ no longer depends
on $r$; $R_K$ must be larger than the optical disc and $C$ is related
to the mass $M$ (bulge and disc) of the galaxy through

\begin{equation}
C \approx {{\left( GM \right)^{1 \over 2}} \over 
{R_F^{1 \over 2} \theta_0}}
\end{equation}

Therefore, even if we are using a set of 4 parameters, observations
very much reduce the choice, which must be different for each galaxy.

Figure 1 illustrates the different rotation curves considered here for
the particular case of the Milky Way. The observational curve has been
taken from Burton et al (1992). Figure 2 is another example for NGC 5023. In
this case, there is no practical difference between the observational
(Bottema et al, 1996) and the assumed disc gas curves.

\begin{figure}
\resizebox{6.5cm}{!}{\includegraphics{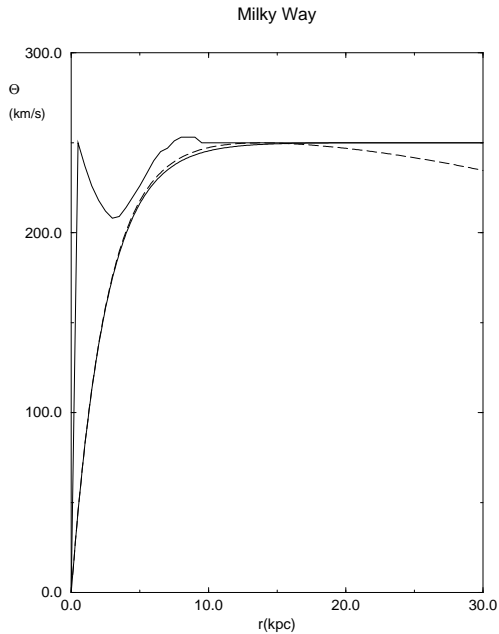}}
\caption[]{ Rotation curves for the Milky Way. The dominant curve at
small radii is observational (Burton et al, 1992). The continuous line
represents the assumed disc gas rotation curve. The dotted line
represents the assumed disc stellar rotation curve. Values of the
parameters: $\theta_0 = 250 km s^{-1}$, $R_F =$ 2.5 kpc, $R_K= 60$
kpc, $C= 2.5$}
\label{}
\end{figure}

\begin{figure}
\resizebox{6.5cm}{!}{\includegraphics{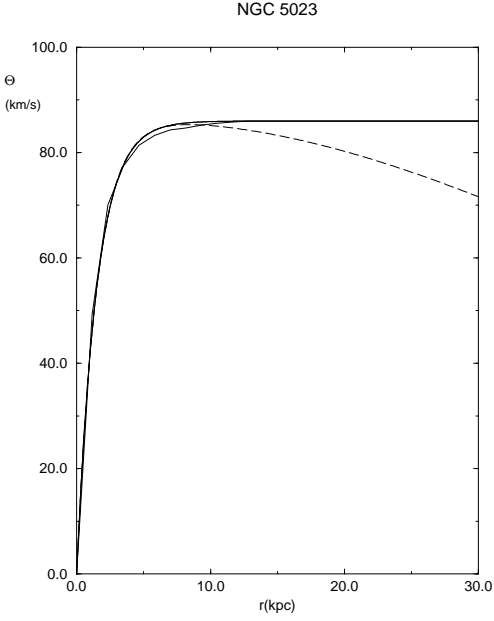}}
\caption[]{Rotation curves for NGC 5023. The assumed disc gas rotation
curve is slightly above the observational rotation curve. The dotted
line again represents the assumed disc stellar rotation curve. Values
of the parameters: $\theta_0 = 86 km s^{-1}$, $R_F =$ 1.5 kpc,
$R_K=50$ kpc, $C=2$}
\label{}
\end{figure}

We further assume an exponential gas distribution (Freeman, 1970), which
is reasonable for large radii, so that the number of stars born at a
given radius throughout the whole history can in turn be assumed to be
exponential

\begin{equation}
\rho_B =A e^{-{r \over R}}=A e^{-\beta x}
\end{equation}

where $A$ is a constant related to the gas-star formation efficiency,
its precise value being unimportant for our present purposes. $R$ is the
radial scale length of the exponential disc and $\beta =R_F /R$ can be
adopted as a parameter, not free because $R$ can be deduced from
observations. The precise definition of $\rho_B$ is: $\rho_B(r)dr$
gives the number of stars born in a ring between $r$ and $r+dr$, throughout
the whole history of the galaxy. It would coincide with the present
distribution of stars if  stars were not able to move to other rings.

The velocity of these $\rho_B(r) dr$ hypothetical stars would be given
by eq. (4) and, as a relation between $r$ and $\theta$ exists, we are
able to calculate $\Gamma (\theta) d\theta$, the number of stars born
with a velocity in the range $[\theta, \theta +d\theta]$

\begin{equation}
\rho_B(r)dr = \Gamma (\theta) d\theta
\end{equation}

\begin{equation}
\Gamma (\theta) ={\rho_B \over {d\theta/dr}}=
A e^{-\beta x} {R_F \over \theta_0} e^{x} =
{{A R_F} \over \theta_0} e^{x\left( 1-\beta \right)} =
{{A R_F} \over \theta_0} \left(1-{\theta \over \theta_0}
\right)^{\beta -1}
\end{equation}

Real stars would conserve the speed they had at birth. Therefore, the
distribution of stars in the velocity space would be the same and
is given by eq. (9). For real stars however, the relation between
$\theta$ and $r$ is different. This relation is now accounted for by
eq. (5). The real distribution of stars in the position space $\rho$
would be obtained by

\begin{equation}
\rho(r) dr = \Gamma (\theta) d\theta
\end{equation}

therefore

\begin{equation}
\rho(r)= \Gamma (\theta) {{d\theta} \over {dr}} =
	{{A R'} \over \theta_0}{\theta_0 \over R'}{{d\varphi}\over{dx}}
	\left( 1-{\theta \over \theta_0}\right)^{\beta
	-1}{{d\varphi}\over {dx}}
\end{equation}

where $d\varphi /dx$ would be calculated using (5) although it is not
necessary to write it explicitly.

Truncation would take place for $\rho(r) =0$, i.e. for $d\varphi /dx
=0$. i.e. the maximum of the function $\varphi(x)$. The equation  $d\varphi /dx
=0$ is transcendent and the value of the truncation radius $r_t$ cannot
be found analytically. We will therefore adopt numerical techniques
for given real spiral galaxies.

Let us consider

\begin{equation}
\mu_B =-2.5 \log{\rho_B \over \rho_C}
\end{equation}

\begin{equation}
\mu =-2.5 \log{\rho \over \rho_C}
\end{equation}

where $\mu_B$ and $\mu$ are in mag arcsec$^{-2}$ and $\rho_C$ is a
constant. Here, $\mu$ is the real stellar luminosity profile and
$\mu_B$ would correspond to the luminosity profile that would be
observed if all born stars had maintained their orbit radius until the
present. In the most internal regions, where magnetic forces are
negligible with respect to gravitational forces, the distribution $\mu_B$
would be the actual luminosity profile. It is easily checked with
eq. (7), that $\mu_B$ depends linearly on $x$ (or on the galactocentric
radius, $r$) defining the assumed exponential profile. However, for
very large $x$ values $\mu_B$ no longer represents the real
profile, as magnetic forces produce a difference between the real
luminosity and the extrapolated exponential luminosity,
i.e. truncation develops. Therefore, let us define the truncation
curve as

\begin{equation}
t(r) = \mu - \mu_B = -2.5 \log{\rho_B \over \rho}
\end{equation}

Therefore

\begin{equation}
t(r)=-2.5 \log{{A\left( 1-{\theta \over \theta_0} \right)^{\beta -1}
{{d\varphi}\over {dx}}} \over {A e^{-\beta x}}}=
2.5 \log{{e^{-\beta x}} \over {\left( 1-\varphi \right)^{\beta
-1}{{d\varphi}\over{dx}}}}
\end{equation}

where again  $d\varphi /dx$ is obtainable from (5).

\section{Results}

Figure 3 represents the truncation curve obtained from the observations and the
theoretical one, obtained with a set of reasonable values of the
different parameters involved, for the galaxies NGC 5023 (fig. 3a),
NGC 891 (fig. 3b) and  NGC 4013 (fig. 3c). The observational
truncation curve has been adopted from van der Kruit \& Searle
(1982a) 
for NGC 5023 and NGC 4013, and  from van der Kruit (1981b)
for NGC 891. For the Milky Way,
no truncation curve is available and we represent in fig. 4 the
theoretical curves for $C= 2.5$ and $C= 3$. To adopt the rotation
curve we have taken observations from Sofue (1996) and Bottema (1996)
for NGC 891 and
NGC 4013 respectively. The values of the radial scale length have been
taken from van der Kruit \& Searle (1982a) for NGC 5023 and NGC 4013,
Porcel et al. (1997) for the Milky Way and van der Kruit \& Searle
(1981) for NGC 891. The truncation radius is $r_t$ defined as
$t(r_t)=\infty$.
For the Milky Way we have estimations of $r_t$ (as mentioned in the
introduction) in reasonable agreement with our theoretical outputs. As
stated above, the involved parameters are only relatively
free. Therefore, we conclude that the agreement between observations
and theory is very good.

\begin{figure*}
\resizebox{12cm}{!}{\includegraphics{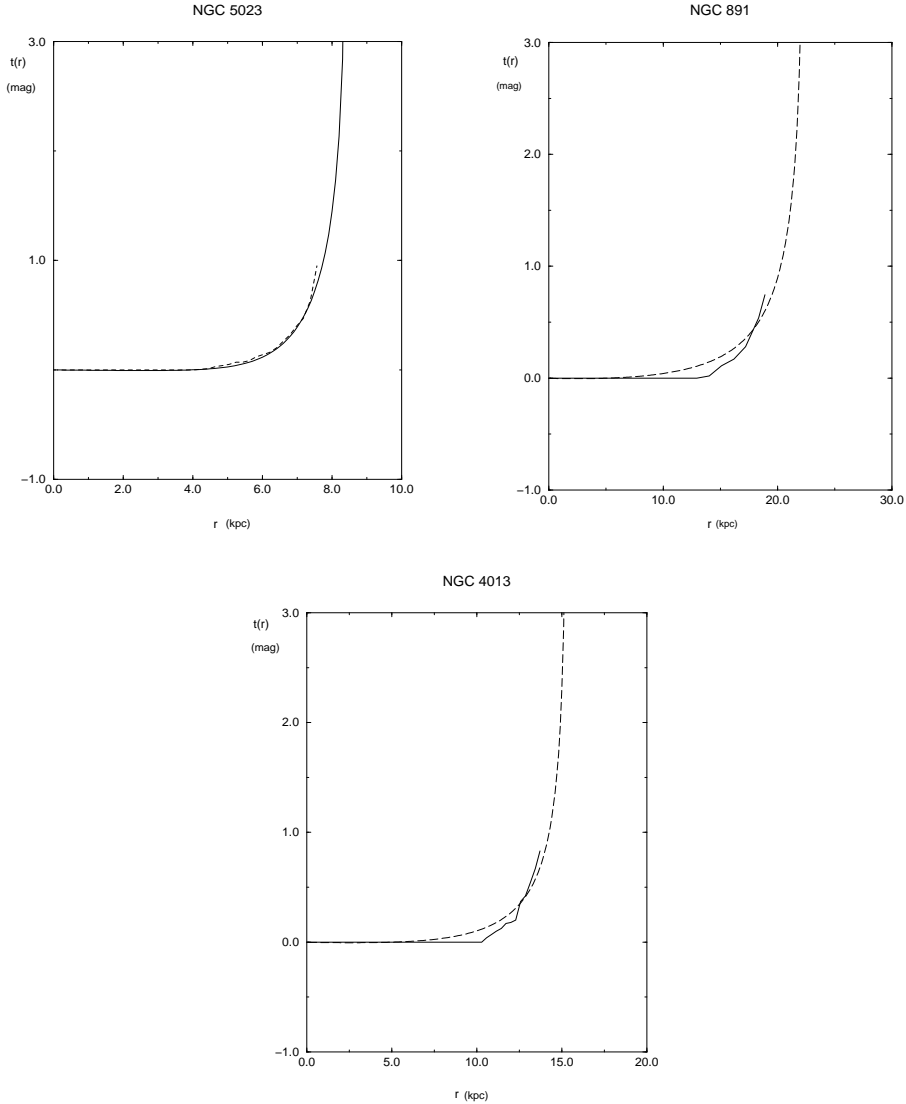}}
\caption[]{Truncation curves of different galaxies. Dotted line,
observational curve; continuous line, this model. Fig. 3a: NGC 5023,
with parameters $R_F$= 1.5 kpc, $R_K$= 50 kpc, $R$= 2 kpc, $C$=2,
$\theta_0 = 86 kms^{-1}$. Fig. 3b: NGC 891, with parameters  $R_F$= 
6 kpc, $R_K$= 75 kpc, $R$= 4.9 kpc, $C$=1, $\theta_0 = 230 kms^{-1}$.
Fig. 3c: NGC 4013, with parameters  $R_F$= 3 kpc, $R_K$= 70 kpc, $R$= 
2.3 kpc, $C$=2, $\theta_0 = 175 km s^{-1}$ }
\label{}
\end{figure*}

\begin{figure}
\resizebox{6.5cm}{!}{\includegraphics{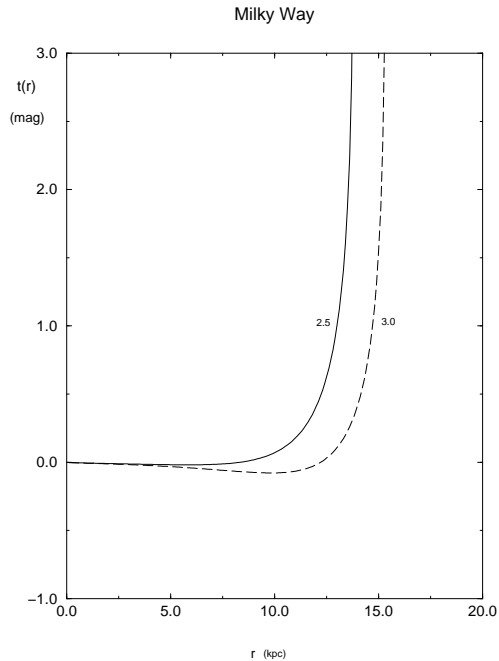}}
\caption[]{Predicted truncation curve for the Milky Way for
$C$= 2.5 and $C$=3. The
remaining parameters were assumed to be: $R_F$= 2.5 kpc, $R_K$= 60
kpc, $R$= 2.5 kpc, $\theta_0 = 250 kms^{-1}$. $R_F$ and $\theta_0$
were adopted taking into account Burton et al (1992). $R$ was adopted from
Porcel et al. (1997)}
\label{}
\end{figure}

Once the validity of the model has been confirmed, we are able to predict  or
compare the dependence of the truncation curve, and in
particular the truncation radius, on the value of the different
observable parameters characterizing the rotation curve.

We predict that the truncation radius is very sensitive to the value
of $R_F$, as shown in Fig. 5. Those galaxies having a rotation curve
slowly reaching the constant rotation velocity (larger $R_F$) would
have a more extended stellar disc (larger $r_t$), with an approximate
relation $r_t \approx 4 R_F$.

\begin{figure}
\resizebox{6.5cm}{!}{\includegraphics{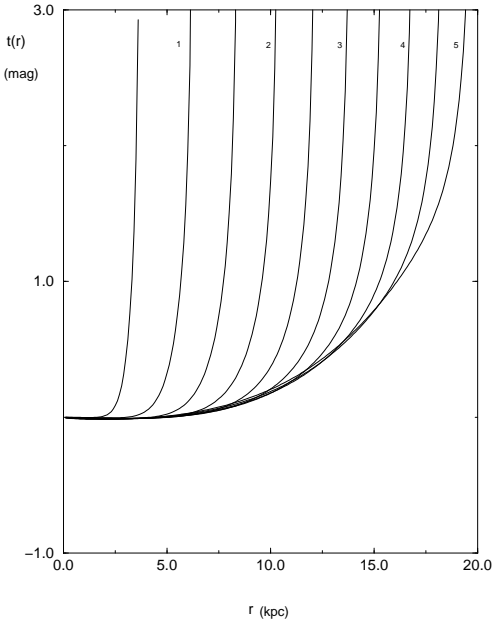}}
\caption[]{Truncation curves for different values of $R_F$. Other
parameters are $R$= 2.3 kpc, $R_K$= 50 kpc, $C$=2 }
\label{}
\end{figure}

As $\varphi(x)$ is independent of $\theta_0$, we see from eq. (14) that
$t(r)$ and $r_t$ do not depend on $\theta_0$. Therefore, our analytical
model predicts no statistical relation of $r_t$ and $\theta_0$, the
rotation velocity at infinity.

\section{Conclusions}

The rotation curve of the gaseous disc determines the truncation of
the stellar disc. We have shown how the rotation scale length of the
gas disc, $R_F$, which represents a typical length for the disc to
reach the outer flat rotation velocity is very closely related to the
truncation radius. However, we find that the truncation radius is
insensitive to the asymptotic rotation velocity at large radii.

To confirm these predictions a larger statistical basis is
needed. Though the sample used by Barteldrees and Dettmar (1994) contained
as many as 27 edge-on galaxies, their rotation properties are mostly unknown.
When these properties are studied by observations,
the low luminosities at which truncation is observed may pose
difficult problems in many galaxies, and hence in the statistical
analysis. At large radii, stars are not observed either because of a
physical truncation or due to sensitivity limitations. In most cases,
truncation takes place at those radii where good photometry is able to
detect it. Nevertheless, large truncation radii could be unobservable because
of sensitivity limits, thus introducing a bias. For instance, when the
radial scale length, $R$, is very small, a high $r_t$ could be
undetectable.

Observations to detect a Keplerian regime of the stellar disc at large
radii may be unsuccessful, as such a region could in
general be devoid of stars or they could possess transient orbits.

The truncation of stellar discs is a fundamental concept in understanding the
evolution and structure of spiral galaxies. Some of the assumptions
adopted here are reasonable, though modifiable, but it can be firmly
concluded that the magnetic scenario explaining flat rotation curves
also provides a clear, simple and natural explanation for this
phenomenon. Other truncation models based on alternative hypotheses
could clarify in the future our understanding of this neglected but
important dynamic feature.

\begin{acknowledgements}This paper has been supported by the spanish ``Ministerio de Educacion
y Cultura'' (PB96-1428) and the ``Plan Andaluz de Investigacion''
(FQM-0108).
 \end{acknowledgements}

\end{document}